\documentclass[conference]{IEEEtran}
\IEEEoverridecommandlockouts
\usepackage{cite}
\usepackage{amsmath,amssymb,amsfonts}
\usepackage{algorithmic}
\usepackage{graphicx}
\usepackage{textcomp}
\usepackage{xcolor}
\usepackage{cleveref}
\usepackage[binary-units]{siunitx}
\usepackage{tabularx}
\usepackage{xspace}

\newcommand{\LDC}{\textit{L1D}\xspace}
\newcommand{\LTWO}{\textit{L2}\xspace}
\newcommand{\LTHREE}{\textit{L3}\xspace}

\usepackage{listings}
\usepackage{ifthen}
\newboolean{ShowAuthors} 
\setboolean{ShowAuthors}{true} 

\def\BibTeX{{\rm B\kern-.05em{\sc i\kern-.025em b}\kern-.08em
    T\kern-.1667em\lower.7ex\hbox{E}\kern-.125emX}}
    
\usepackage{titlesec}

\begin{document}

\title{Online Model Swapping in Architectural Simulation
}
\ifthenelse{\boolean{ShowAuthors}}
{
  \author{\IEEEauthorblockN{Patrick Lavin, Jeffrey Young, Rich Vuduc}
  \IEEEauthorblockA{\textit{Computational Science and Engineering} \\
  \textit{Georgia Institute of Technology}\\
  Atlanta, GA \\
  {plavin3,jyoung9,rvuduc3}@gatech.edu}
  \and
  \IEEEauthorblockN{Jonathan Beard}
  \IEEEauthorblockA{\textit{Arm Research} \\
  Austin, TX \\
  jonathan.beard@arm.com}
  }
}{
  \author{\IEEEauthorblockN{Author names withheld for blind review.}}
}

\maketitle

\begin{abstract}
As systems and applications grow more complex, detailed simulation takes an ever increasing amount of time. The prospect of increased simulation time resulting in slower design iteration 
forces architects to use simpler models, such as spreadsheets, when 
they want to iterate quickly on a design. However, the task of migrating from a simple simulation to one with 
more detail often requires multiple executions to find where 
simple models could be effective, which could be more expensive 
than running the detailed model in the first place. Also, architects must often rely on intuition to choose these simpler models, further complicating the problem.

In this work, we present a method of bridging the gap between simple and 
detailed simulation by monitoring simulation behavior online and automatically 
swapping out detailed models with simpler statistical approximations. 
We demonstrate the potential of our methodology  
by implementing it in the open-source simulator \textit{SVE-Cachesim} to swap out the level one data cache (\LDC) within a memory hierarchy.
This proof of concept demonstrates that our technique can handle a non-trivial use-case in
not just approximation of local time-invariant statistics, but also those 
that vary with time (e.g. the \LDC is a form of a time-series function),
and downstream side-effects (e.g. the \LDC filters accesses for the level two cache). 
Our simulation swaps out the built-in cache model with only an $8\%$ error 
in the simulated cycle count while using the approximated cache models for 
over $90\%$ of the simulation, and our simpler models require two to eight times
less computation per ``execution'' of the model. 
\end{abstract}


\section{Introduction}\label{sec:intro}
With traditional lithography-scaling
slowing and Dennard scaling effectively over~\cite{moore1965cramming,chien2013moore,dennard1974design}, performance increases will come 
increasingly from specialization and scaling ever upward
and outward. That specialization 
will be in both the compute and memory/storage technology
domains. Rapid prototyping tools that enable system 
architects to determine the best composition of 
algorithm, architecture, and scale are critical to enabling 
the development of next-generation systems. Computer architecture
modeling today
is largely accomplished through discrete event simulation, 
therefore the time to model a system is roughly proportional
to the number and detail of the components modeled.
To assess a given architectural configuration, modelers set up
a discrete event simulation with a 
fixed architectural topology. Each model within this topology is 
typically fixed within 
the simulation infrastructure for the duration of simulation.
However, applications are not fixed, and they typically have rich variation in 
activity from one point in time to the next, with some phases
of relatively stable behavior. Specifically, we intend to dynamically
swap from complex to simple models when possible for these phases of stable behavior. Detecting when, where, and how to select simpler models while ensuring equivalent model fidelity is the primary contribution of this work.

\begin{figure}[htp]
    \centering
    \includegraphics[width=\linewidth]{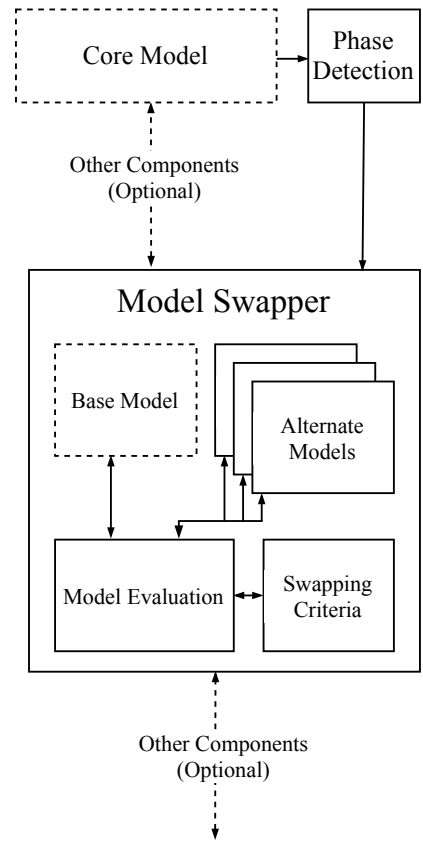}
    \caption{A diagram of the model swapping components. Dashed lines represent components that should already exist in a simulation and solid lines represent components covered in this work. In this diagram, \textit{Other Components} means other simulation components that will remain intact and not be swapped out.}
    \label{fig:modelswapping}
\end{figure}

Computer architecture simulation (simply \textit{simulation} 
from here on out for brevity) is a critical tool
for determining how to shape future architecture. \textit{Simulation} is largely
synonymous with discrete event simulation of a computer
system-on-chip~\cite{akram2019survey}, with each sub-component 
of the simulator representing some module within the target architecture.
Abstractly, each module is some mathematical function that
maps input features of a time series to some output response.
Early \textit{simulators}, such as SimpleScalar~\cite{austin2002simplescalar},
modeled every function or component of the system in some level of 
detail. This practice was fine for the age in which they lived,
however simulation of a modern multi-core system-on-chip with a 
framework such as SimpleScalar would take an inordinate 
amount of time. More recent simulators such as gem5~\cite{binkert2011gem5}
adopt a method known as ``sim-points''~\cite{sherwood2002automatically}
in order to speed simulation by adopting a statistical approximation
for the behavior of entire phases across an entire simulation. 
While this approach speeds simulation, it bypasses many components
that architects would want to model in detail, and it assumes that the component model for the entire 
system for a given phase is the same. This works well for the entire simulation, but perhaps we want
to speed most of the simulation but model one specific component 
of interest in detail; that is where our \textit{online} modeling
approach applies. Our approach enables
online migration to simpler models and faster execution with a single
run of the simulator, as opposed to multiple runs with ``sim-points''. Our proof of concept shows that it is possible 
to swap in simpler, more abstract models while minimizing the
impact to the rest of the system. 

Modern computer architects have a myriad of computer simulation
frameworks to choose from (e.g.~\cite{binkert2011gem5,marssx86,sst,
heirman2012sniper}). This work will apply to many of them, assuming that
 simulator framework is ``plug-able'' (e.g.~\cite{sst}),
that is, the simulator modules themselves have a defined interface. 
We assume that the plugin mechanism defines any state transfer 
between plugin modules and that state transfer only occurs at
module boundaries through defined interfaces or ports (i.e., no global state).
Within this framework, model swapping is trivially possible, that is
the reader can surely imagine changing out components by re-connecting
input and output ports (i.e. Figure~\ref{fig:modelswapping}. 
This oversimplification glosses over 
necessary breakage of this pure model (e.g. synchronization primitives
must exist, and often have state), however, it is enough to build on. 
With this definition of a plug-able framework, we next need to decide how to swap modules while respecting the spatial and temporal output that downstream models expect (e.g. the level one data 
cache filters for the level two data cache, and therefore the behavior
of the upstream component necessarily impacts the downstream one). 
Our work examines swapping in a model that is specifically chosen because its behavior has these cascading effects. 

Swapping a complex model for a simple one could make a relatively large
improvement on the time needed to execute a simulation. As 
an example, moving from a component that requires eight memory operations
and two comparisons to a component that only needs a single comparison could
greatly impact simulation efficiency, assuming that component is 
utilized extensively. For many applications~\cite{nai2015graphbig} 
the ratio of memory operations to compute is high, therefore the memory
subsystem is on the critical path, being heavily utilized for much of 
the simulation. Swapping to simpler models on this critical path, such as the level
one data cache could have a huge impact.
Our primary contributions, 
to be detailed within this work, are summarized by the following: 1) We present a proof of concept for online model swapping within an architectural
simulation. 2) We demonstrate multiple statistical models that can approximate a functional cycle-accurate model 3) We detail a model selection methodology that can be used for online model selection.
\section{Online Model Swapping}\label{sec:onlineswapping}

\begin{figure}[htp]
    \centering
    \includegraphics[width=.7\linewidth]{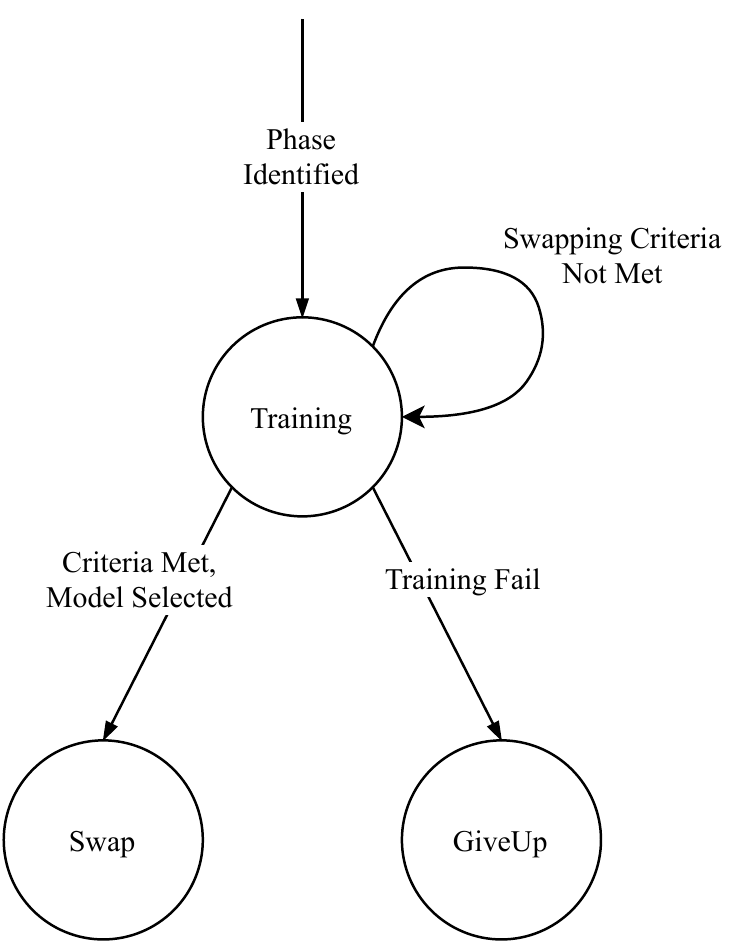}
    \caption{A state diagram for the abstract model swapping algorithm. State transitions happen upon interval end, as defined by the Phase Detector. Alternate models are trained per-phase, and once swapping criteria have been met, that phase will move into the \textit{Swap} state, where a single alternate model is chosen for that phase for the rest of the simulation. If training seems too hard for a phase, or takes too long, a \textit{GiveUp} state can be entered, so that we can stop spending computational resources on a phase deemed too complex.}
    \label{fig:swapalgo}
\end{figure}

The scope of an idea like model swapping is quite broad and needs to be narrowed down before we can make headway with an implementation. Let us first take a high-level look at the four components of our scheme, which, in general, are answers to the four following challenges:

\begin{itemize}
    \item \textbf{How do we find sections of execution that are simple enough to model?} Our goal of replacing components of a simulation with easy-to-compute statistical models relies on us identifying phases of execution for which our components behave in a way that we predict will be represented well by the models available to us. This will be addressed by \textit{Phase Detection}.

    \item \textbf{What do we swap in place of the base, detailed model?} We need to choose statistical models to provide to our model swapping algorithm. Criteria for such models will be addressed in \textit{Alternative Models}.

    \item \textbf{How do we evaluate alternative models?} As we will have multiple statistical models available to us at simulation time, we must identify any criteria that we think will be useful for ranking the relative performance of the models. Things like model speed and performance must be compared. The question of evaluating models will be addressed in \textit{Model Evaluation}.

    \item \textbf{How do we decide to swap out a model?} At simulation time, we need criteria for deciding when it will be beneficial to swap out the base model for one we have trained ourselves. Deciding to swap the models means considering the criteria previously mentioned and decided if it will be worth it to make the swap. Making the decision between models is covered in \textit{Model Swapping}.

\end{itemize}

\noindent
We will answer these four questions in the following subsections.

\subsection{Phase Detection}\label{subsec:phasedetect}

In order to train simple statistical models to predict behavior, we need to break up the computation into chunks which themselves exhibit simple behavior, at least for the component we want to swap out. This can be accomplished with the help of phase detection\cite{Hind2003PhaseSD}, which, as the name implies, detects phases within a program. Phases represent portions of the program that display similar characteristics, such  as having a consistent number or branch misses or the same set of instruction pointers. Many algorithms exist in this field, but we will only use one in this work, which is laid out in \Cref{poc:pd}. Abstractly, it is important for whatever phase detector that is chosen to be able to identify a phase identifier (\textit{ID}) for the most recent interval, and for it to share that \textit{ID} with the other components of the simulation, so that they know what phase has just run, and can use that info to pick models.

The phase detector communicates phase \textit{ID}s to the other parts of the system. It acts as an event notifier that wakes up the model evaluation function periodically so that it can train and score models.

\subsection{Alternative Models}\label{subsec:alternativemodels}

An obvious need for model swapping is to have multiple models to choose from. The only requirement for these models is that they statistically approximate the original base model responses given a time series of inputs. For instance, a cache model that predicts a miss at time $T$ sends a response to a downstream cache level (e.g. the level two data cache). If the model receives any sort of coherence traffic that needs a response, the model must support this too.
\subsection{Model Evaluation}
\label{sec:method:modelselection}

Various criteria exist for choosing between different statistical models based on things like model size and the number of required sample points. For this work, we highlight the four parameters that need to be considered in any model selection methodology:

\begin{itemize}
    \item \textit{Accuracy} - the statistical model should accurately represent the behavior of the base model it is replacing,
    \item \textit{Side Effects} - the statistical model should not disturb the statistics of other parts of the simulation,
    \item \textit{Model Size} - the model should ideally be smaller than the base model
    given that memory access is likely dominant in the simulation,
    \item \textit{Model Complexity} - the statistical models should ideally require less time to perform prediction than the base model.
\end{itemize}
These four criteria will enable us to limit the choice of models thereby giving us a method of choosing one online during 
simulation.

\subsection{Model Swapping}\label{subsec:modelswapping}

The final part part of any model swapping is the algorithm that decides when it is time to stop training a statistical model and to swap one in. This process is depicted in a state diagram in \Cref{fig:swapalgo}. This algorithm is run per-phase, and the transitions between states occur on interval boundaries. 

To instantiate such an algorithm, all of the other components described so far must be in place, and swapping criteria must be defined so that this algorithm can eventually terminate. Optionally, a time limit can be set so that training is permitted to fail, at which point the swapping algorithm gives up on trying to train a model for that particular phase. 

\section{Proof of Concept: Swapping the L1 Data Cache}
Now that we've explained all the components of our model swapping methodology, it's time for us to implement it. To accomplish this, we implement model swapping for the L1 data cache (\LDC). We choose the \LDC specifically 
because it 
requires non-trivial models to predict behavior,
and changes to this model have real consequences 
for downstream models consuming its output. In this section, we'll discuss the phase detection, alternate models, and model swapping algorithm that we have chosen, and in \Cref{sec:results}, we'll discuss how well our method works.

\subsection{Phase Detection}
\label{poc:pd}
In order to reduce the degrees of freedom for our proof of concept, we opt for a simple solution to the phase detection problem, based on working sets, which was first described in \cite{phase_analysis}. We define an interval as a continuous block of 10,000 instructions. In our case, instructions consist of only memory references, as this is the information that will be available to the \LDC. For each interval, the instruction pointers are hashed into a bit-vector, which serves as the signature for that interval. We can compare interval signatures with a simple similarity metric, and if we have enough consecutive similar intervals, we will classify this as a phase, and begin training models to fit that phase. 

We have included the parameters (such as the signature size and minimum number of similar signatures required to declare a new phase) as well as pseudo-code for this phase detection algorithm in Appendix~\ref{appendix:phase}. While simple, this algorithm faithfully reproduces the known phases for the program. Thus, we deem it suitable for our proof of concept. 

\subsection{Alternate Cache Models}
We need a selection of models to use in place of the \LDC cache when we swap out the base model. In this work, we present three simple models. As we are working to replace a cache, the behavior we need to replace is that of the \textit{hit check} where we ask a cache whether or not it has a line in residence. Thus, a cache model needs to take information about a request (such as the instruction pointer and whether it is a read or a write) and use this to predict whether the access is a hit or a miss.

\begin{figure}[htp]
    \centering
    \includegraphics[width=.6\linewidth]{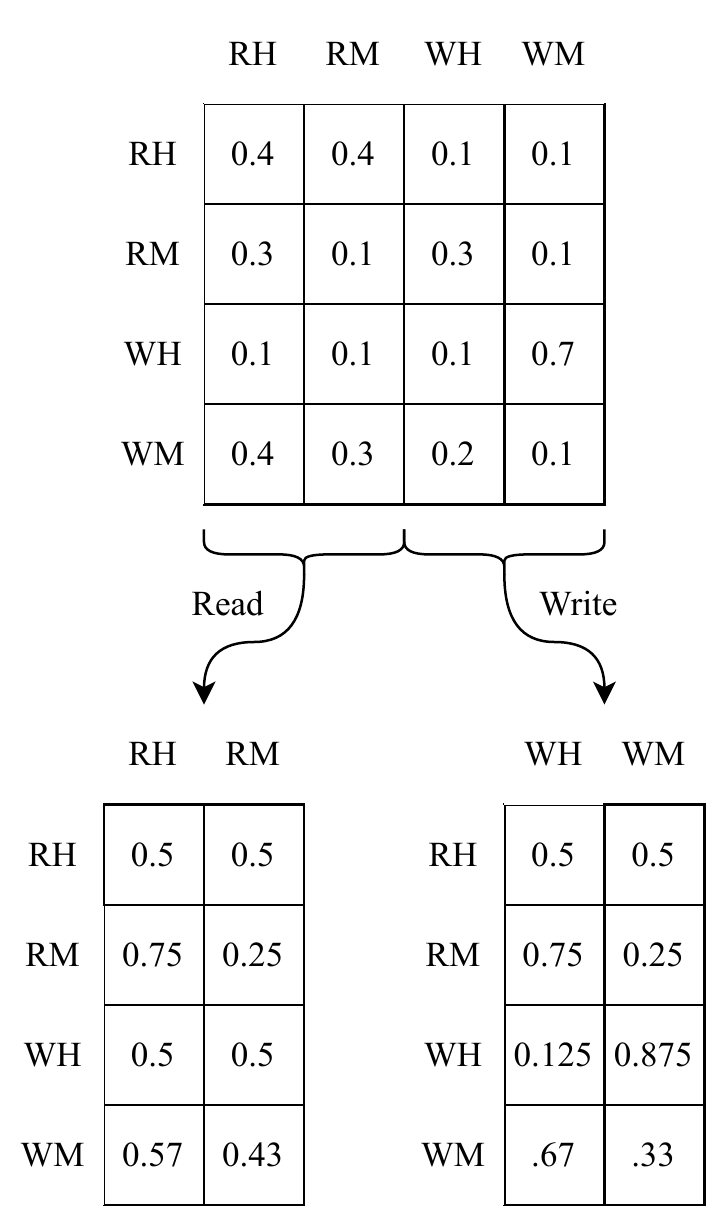}
    \caption{\textbf{Markov-Chain Based Models}: Our Markov-chain based models restrict the transition probabilities based on information regarding the next access. If the next access is a read, we will only allow the model to predict a transition to a \textit{ReadHit (RH)} or \textit{ReadMiss (RM)} state. This image depicts the 4-state Markov model, but the 8-state model works similarly.}
    \label{fig:markov_desc}
\end{figure}

\subsubsection{Fixed Hit Rate}
The first and most basic model is the \textit{Fixed Hit Rate Cache Model}, also referred to as the \textit{Fixed Rate Cache Model}. In this model, we count the number of hits and misses during a phase's execution, and calculate a hit rate for the phase. When using this model for prediction, we generate a random number uniformly spaced between zero and one, and check whether that number is below the hit rate or not. If it is, we say the access is a hit, otherwise, we classify it as a miss, so we must communicate this to the \LTWO, so as to disturb the rest of the simulation as little as possible.

\subsubsection{4-State Markov Model}
\label{sssec:markov}
The fixed rate model is simple, but it is quite likely that it will not be able to accurately model most workloads, as it has very little state to approximate true application behavior. To add some history (and therefore state) to our prediction, we use a Markov-chain based model. Our first Markov model includes four states, \textit{ReadHit, ReadMiss, WriteHit}, and \textit{WriteMiss}. During training, we learn the transition probabilities from one state to the next. During prediction, we generate a uniform random number and move to the state indicated by the transition probability. However, we have one caveat: if the next request to the cache is a read, we do not want our Markov model to predict that we move to a \textit{WriteHit} or \textit{WriteMiss} state. Thus, during prediction, we restrict the model to moving to states corresponding to the current cache access. This behavior is depicted in \Cref{fig:markov_desc}, and the pseudo-code is shown in Appendix~\ref{appendix:markov}. 

\subsubsection{8-State Markov Model}
While the 4-State Markov model adds a single piece of history, neither it nor the Fixed Hit Rate model capture one of the most important performance features of a cache, namely spatial  and temporal locality. To remedy this, we have an 8-state Markov-chain model, that adds a small amount of locality information. It has the same states as the above 4-state Markov model, but it has both \textit{Near} and \textit{Far} versions of each of the four states. A \textit{Near} state is defined as cache access occurring on the same \SI{64}{\byte} cache line as the last one, and a \textit{Far} access as anything else. Similarly to the 4-state model, we must restrict the states we allow the model to transition to during prediction based on both whether or not the next access is a read or a write, and also whether the access is considered  \textit{Near} or \textit{Far}. This means that prediction is only able to transition to one of two states, so the entire prediction takes only two comparisons (one to determine if the access is near or far, and one to determine if it is a hit or a miss).

\begin{table}
\caption{Model Complexity}\label{tab:complexity}
\begin{tabularx}{\linewidth}{|l|X|X|X|}
\hline{}
 & Size (Bytes) & Hit Check (comparisons) & Training (comparisons)  \\ \hline\hline
Base & 8192 & 16  & 16\\ \hline
Fixed Rate & 16 & 1 & 1\\ \hline
Markov 4 & 384 & 1 & $O(N^2)$ \\ \hline
Markov 8 & 1536 & 2 & $O(N^2)$\\ \hline
\end{tabularx}
\end{table}

\subsection{Model Evaluation Criteria}
\label{ssec:modeleval}
As mentioned in \cref{sec:method:modelselection}, we have identified four areas to evaluate models on. For a cache component, we have defined them as follows:

\subsubsection{Accuracy}
For our purposes, accuracy is defined as the percentage of hits correctly predicted. This is accomplished during simulation by training the statistical models as soon as we have any data on the phase (i.e., as soon as it is identified by the phase analysis component), and running the partially trained models alongside the base model until we have a good idea of the accuracy. 

\subsubsection{Side Effects}
In our case, the most important side effect to consider is the locality of the references that miss in the \LDC, as these are the ones that go to the \LTWO. Many metrics for locality are computationally intensive, and since our goal is to speed up simulation, we have chosen a simple proxy for locality, the \textit{near miss count}. With \textit{near} being defined as in the 8-state Markov model: if two accesses are on the same \SI{64}{\byte} cache line, they are considered near, and otherwise classified as far. Thus, the metric is simply the number of misses that are classified as near. We can compare this with the number of near misses from the base model to get an idea of the locality properties of the accesses going to \LTWO. As with accuracy, we will need to run this model alongside the base model during training to get this number. 

\subsubsection{Model Size}
Aside from metrics derived at runtime, there are some static information we can use in model selection, the first of which is the model size. 
\begin{itemize}
\item The base model is a \SI{32}{\kibi \byte}, 8-way set associative cache, with \SI{4}{\byte} words. This gives us 128 sets with 8 tags each, and stored as \SI{8}{\byte} integers. This is a total size of \SI{8}{\kibi \byte}.

\item The Fixed Hit Rate model requires storing an \SI{8}{\byte} hit count, and an \SI{8}{\byte} hit rate, meaning the full size is only \SI{16}{\byte}. 

\item The Markov models each require storing an $N\times N$ transition count matrix, and an $N\times N$ transition probability matrix, where $N$ is the number of states. They also memoize the restricted transition matrices (depicted in \Cref{fig:markov_desc}. This amounts to another $\frac{N}{2}$ matrices of size $N\times2 $ which brings the total cost to $3N\times N$. Assuming that these are all stored with 8-byte values, we end up with \SI{384}{\byte} for the 4-state Markov model, and \SI{1536}{\byte} for the 8-state model. 

\end{itemize}

\subsubsection{Model Complexity}
We measure model complexity as the number of comparisons required to make a hit check. For the base model, this takes 16 comparisons, as it does a linear search over the tags present in the set, which is 8-way associative. Another search will be needed for eviction. For the Fixed Hit Rate Model, a single comparison is needed to see if the random number generated is less than the hit rate. 

Due to the way the Markov models restrict the states they can move to, we actually only need a single comparison for the 4-state model, which the 8-state model needs an additional comparison, to check if the current address is near or far. This was explained in depth in \Cref{sssec:markov}. The numbers for the model size and complexity are also listed in \Cref{tab:complexity}. 

\subsubsection{Model Score}

We compile the above into a length-4 vector representing the model's performance on the phase. The first number is the model accuracy, which of course is a number between zero and one. The second value is count of near misses predicted by the statistical model, divided by the number of near misses that the base model emitted. The third number is the size of the model as a fraction of base model, and the final number is the number of comparisons required as a fraction of the base model. Naturally, the ideal vector for a model would be $[1,1,0,0]$, as this would represent perfect accuracy, getting the number of near misses exactly correct, as well as having zero size and zero complexity. While a metric such as cosine similarity could give us a ranking, we opted for a method in which the magnitude of the vectors mattered. Thus, we assign each model a score which is the L2-norm of the difference between their score vector and the ideal vector, which means that a lower score will be better. These scores will be used by the model swapping algorithm in the next section to choose between models.

\subsection{Model Swapping}
\label{sss:model_swapping}

In \Cref{fig:modelswapping}, we showed an abstract picture of what model training and swapping should look like. For this work, due to the simplicity of the models chosen, we have used a reduced version of that diagram for our algorithm. We use a single training criteria, which is simply a check on whether we have trained for two intervals or not, and we have no \textit{GiveUp} state. Thus, we always swap in the best scoring model after training for two intervals, regardless of any other factors. We found that this time period was long enough for the learned parameters in our models to stabilize, and that the resulting accuracy was reasonable for the total run. 

Now that we've seen the components of our simulation, it is time we take a look at how well this methodology works. In our tests, the time spent with the base model is five intervals per phase to identify, plus two intervals per phase to train, plus the number of any uncategorized phases. This means that for the results presented in \Cref{sec:results}, the swapped models are used for over $90\%$ of the simulation.
\section{Results}\label{sec:results}

\subsection{Methodology} 

\subsubsection{Simulator}
Our technique for model swapping was implemented in \textit{SVE-Cachesim}, which is an in-order, Python-based cache simulator developed for \cite{asvie}. While a simple model such as this ignores some parts of the cache system, such as the effects of pre-fetching, by focusing on a component in the middle of the system, the level one data cache, we were able to study cascading downstream effects introduced by model swapping, and pick models that minimized this disruption. 

\subsubsection{Trace}

To test our phase-analysis methodology, we used a micro-benchmark called Meabo\footnote{https://github.com/ARM-software/meabo}. Meabo allows us to run a program with pre-identified phases, which makes it easy for us to visually confirm that our phase analysis is working. We chose three phases from Meabo, \textit{Phase 1: Floating-point \& integer computations with good data locality}, \textit{Phase 4: Vector addition}, and \textit{Phase 10: Random memory accesses}. These were chosen as we expected them to vary in their utilization of the cache hierarchy. 

We made a few changes to the code, which are intended as markers to ensure that we have correctly identified each specific phase. First, we added a loop to Meabo so that all phases will be run multiple times instead of just once, and second, we added a marker phase between each phase so we can be sure which phases are from Meabo and which are belong to non-execution artifacts, such as data initialization. 

The trace is collected by running Meabo with a single 
thread on an Intel Xeon Haswell E52690-v3 and capturing all memory references (from virtual address space)
with DynamoRIO \textit{memtrace} 
\cite{dynamorio}. The trace is roughly three million 
references long.

\subsubsection{Model Training}
As mentioned in \Cref{sss:model_swapping}, we use a simple model swapping criteria in this paper: one or more models are trained for the first two runs of an interval, and then the best one is selected. From inspection, we saw that the training parameters for all of our selected models stabilized quickly, and thus we did not need to train for more time. Additionally, we do not use a criteria to declare training a failure in this work. 

\begin{figure}[htp]
    \centering
    \includegraphics[width=\linewidth]{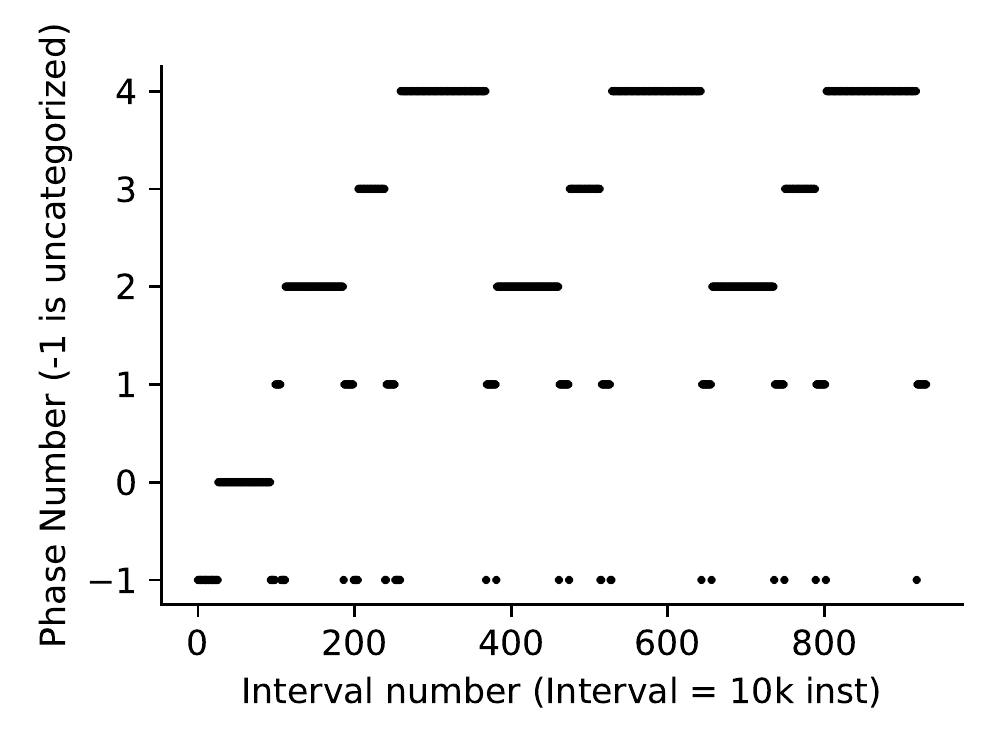}
    \caption{\textbf{Meabo phases}: Here we have plotted the interval number of the program along with its corresponding identifier assigned to it through phase detection. These images shows our marker phase, phase one, being run after every Meabo computational phase, phases two, three, and four.}
    \label{fig:meabo}
\end{figure}

\subsection{Phase Analysis}
First, we should examine that our phase analysis code is working as intended. In \Cref{fig:meabo}, we plot the phase identifier (\textit{id}) for every interval in the simulation. At the end of every interval, which are each 10,000 instructions, the phase analysis code will attempt to identify the interval by comparing the phase signature to those it has encountered previously. If it is successful an \textit{id} is assigned, otherwise the interval is labeled as (-1). 

The trace we collected ran each phase three times. Looking at \Cref{fig:meabo}, it is clear then that phases two, three, and four must be the phases from Meabo, and that phase one must be our marker phase that runs between Meabo computational phases. For the rest of the paper, we will refer to these phases by the number assigned to them by our phase detection algorithm, but we will include a description of the computation (e.g. \textbf{\textit{High Locality}}, \textbf{\textit{Vector Add}}, \textbf{\textit{Random}}, \textit{etc.}) so that it will be clear to which we refer.  

Now that we've demonstrated empirically that our phase detection methodology is sound, and capable of driving our online model selection process. We can now evaluate how accurate each model is.

\begin{figure}[htp]
    \centering
    \includegraphics[width=\linewidth]{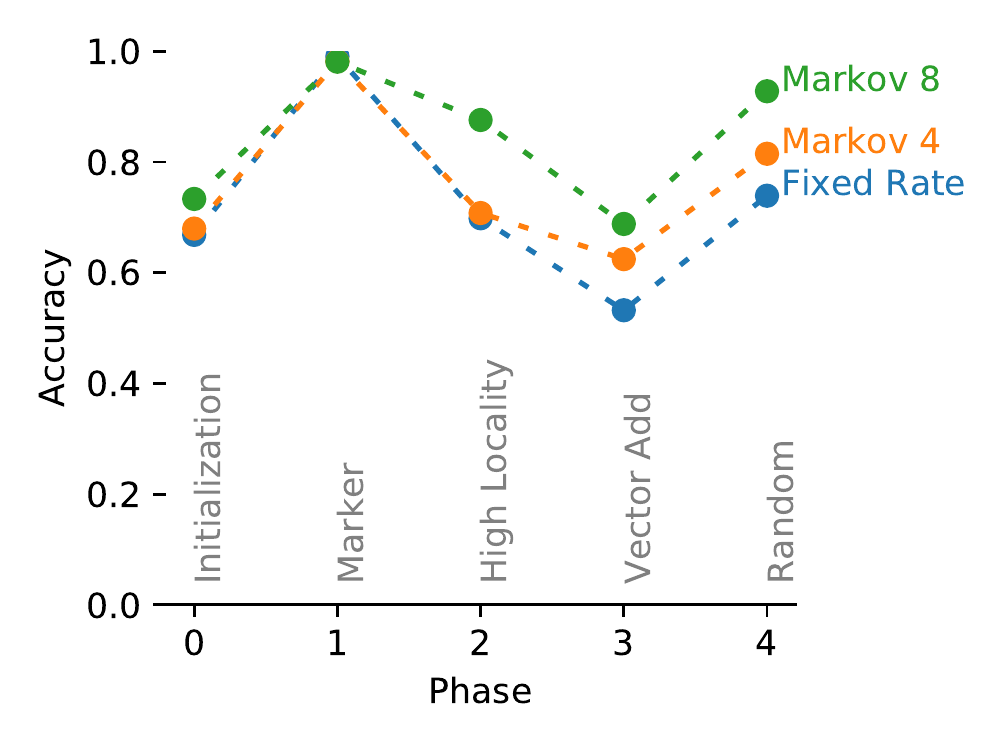}
    \caption{\textbf{Accuracy per phase}: We run our simulation but restrict it to using a single alternate model (as opposed to letting the simulation choose between them), so that we can view their relative performance. We average the accuracy over 10 runs. These values are also in \Cref{tab:acc}, along with the standard deviation in \Cref{tab:stddev}, which was too small to plot.} 
    \label{fig:acc}
\end{figure}

\subsection{Prediction Accuracy}
We would like to compare the accuracy of the each model we have chosen so that we can get an idea of how they measure up against each other. In \Cref{fig:acc}, we measure the accuracy of each model for each phase. Accuracy is defined as the number of properly classified accesses, which means the number of accesses correctly predicted to be hits or misses, divided by the total number of accesses to the level one data cache. We consider the truth to be the standard detailed level one cache model running with no model swapping. 

This plot shows us that the 8-State Markov model is the best for all but the marker phase, Phase 1. The marker phase is essentially all read hits, and thus trivial to predict. The gaps between the model performance on the Meabo phases (Phases 2, 3, and 4) give us hints as to what type of information is important for modeling each phase. In phase 2, the tiny gap between Markov 4 and the Fixed Hit Rate model shows us that knowledge of whether the access was read or write (and whether the last access was read or write) is not important, at least not on its own. This information, however, is useful in both Phases 3 and 4. As we see from the performance of the 8-state Markov model, even the relatively small amount of locality information gained from classifying accesses based on whether they share a cache line with the last access is enough to achieve 90\% accuracy in Phases 2 and 4. 

While promising, this data is an aggregate over the entire run, with multiple runs of phases aggregated into very few data points. We will break this data down to examine it further. 

\begin{figure*}
  \centering\includegraphics[width=1\textwidth]{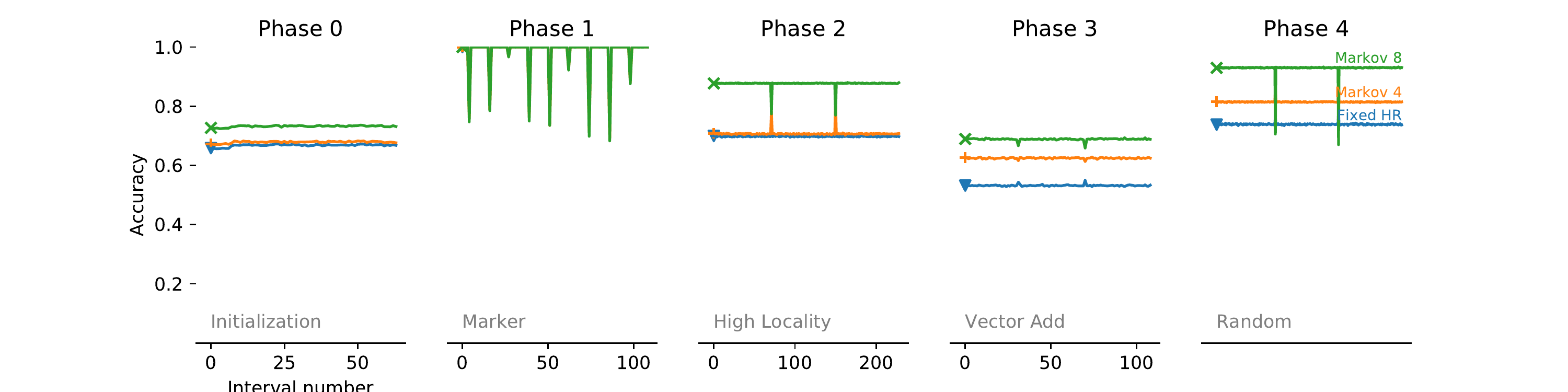}
  \caption{\textbf{Accuracy per interval}: We can view the accuracy that each model achieves on each interval of the trace. This shows us that accuracy does not deteriorate over time, but that there are dips in accuracy occurring upon phase re-entry. For example, Phase 2 has two reentry points (as shown in \Cref{fig:meabo}), and we see corresponding dips when accuracy is plotted sequentially here}
  \label{fig:acc_over_time}
\end{figure*}

\subsection{Prediction Accuracy Over Time}

While it is useful to look at data aggregated per-phase, it leaves us unable to ask question about how the accuracy changes over time, particularly at phase boundaries. \Cref{fig:acc_over_time} displays the accuracy per-interval, as opposed to the per-phase data we looked at the in the last section. 

From the data for Phases 1, 2, and 4, it is clear that the accuracy does change over time. If we reference the phase \textit{id} plot in \Cref{fig:meabo}, we can see that these drops in accuracy occur when phases are re-entered. This is perhaps the opposite of the expected behavior - we would expect accuracy to drop at the end of a phase, because the phase detector will not be able to tell the model swapping algorithm that a phase has ended until it identifies an interval that is not a part of the phase. Thus that one interval will have been run with a model trained for a different phase. However, as we only see 3 spikes on Phase 2, it must be that this is happening on phase re-entry, not phase exit. It is likely that phases exhibit different memory behavior upon re-entry, such as a higher number of cache misses. This behavior is likely not captured by our models that assume the behavior has reached a steady state. 

On the other hand, it is quite promising that these lines appear to be flat. This means that the accuracy is not getting worse as time passes. In other words, deleterious state doesn't accumulate or compound over time. 

Now that we have looked at how the models work in regards to the \LDC statistics, we should take a look at how the rest of the simulation is affected. 

\subsection{Locality of Misses}

\begin{figure*}
  \includegraphics[width=1\textwidth]{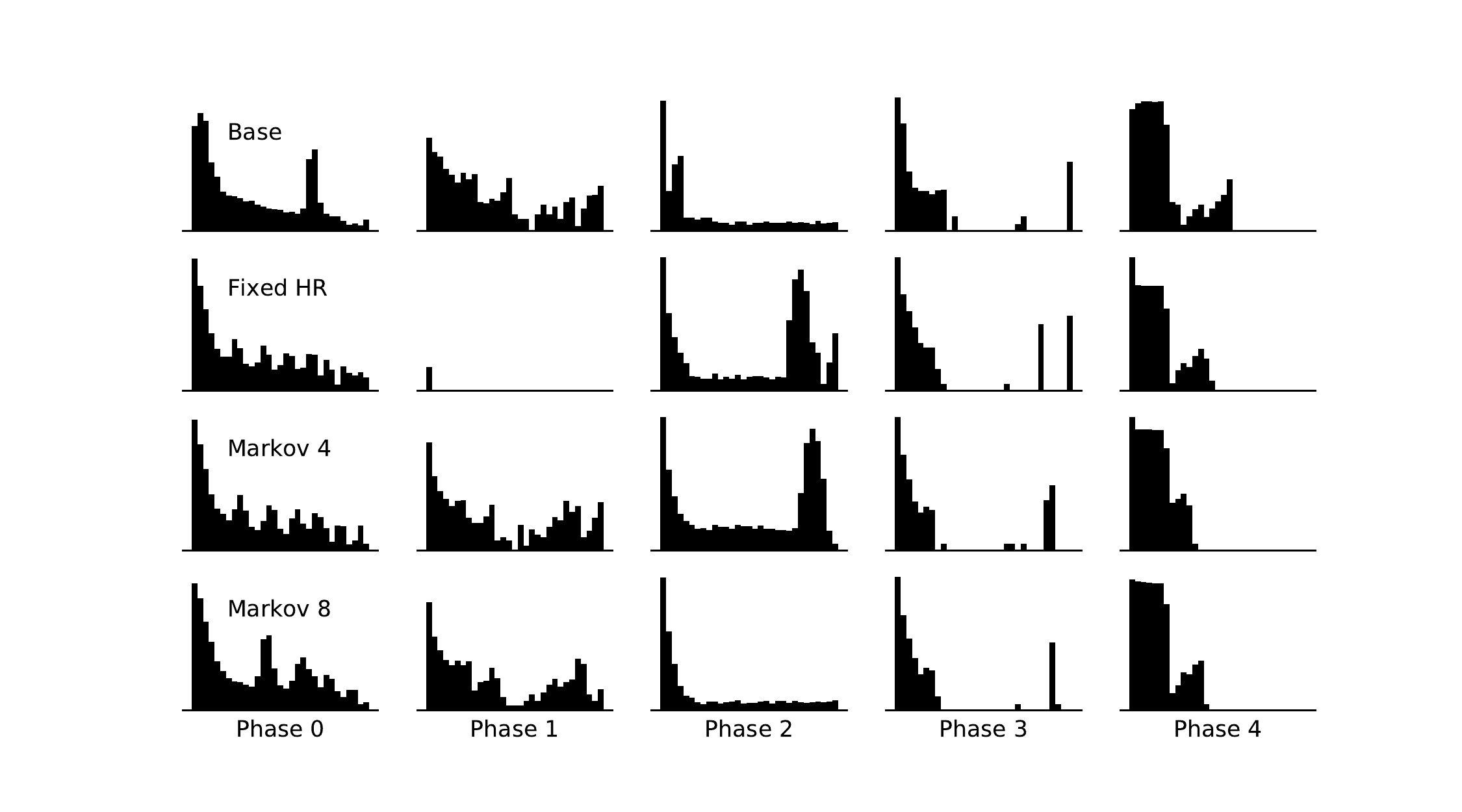}
  \caption{\textbf{Reuse distance histograms}: We calculate the reuse distance for every access to the \LTWO cache and plot this for a single run of the simulator. We cut the x-axis at reuse distance = 500 and the y-axis at count = 125,000. Log y-axis. This shows us that the 8-state Markov model is able to model the locality properties of the base cache much better than the other two statistical models.}
  \label{fig:loc}
\end{figure*}

In \Cref{fig:loc} we have plotted the reuse distance histograms for the cache accesses that miss in the \LDC cache and thus go to the \LTWO cache, which has not been swapped out for a simpler model. It will be important that these match the base model so that we do not disturb the \LTWO behavior too much. Across the top of the plot is the reuse data for the base cache. These are the shapes that we would like our statistical models to learn. 

The locality of phases 0 and 1, program initialization and the marker phase, seem difficult to capture with our models. Phase 0 has a large spike that no models capture, and phase 1 is particularly hard for the Fixed Rate Model. Thankfully, these do not take up a large percentage of the simulation. Moving to the Meabo phases, we see that the 8-State Markov model shows a strong ability to mimic the reuse distances of the base model, especially compared to the other two models. In Phase 2, both the Fixed Hit Rate and the 4-State Markov produce large spikes in the histogram that should not be there, whereas the 8-State Markov model does not. In Phase 3,  the 8-State model seems to do best at reproducing the right side of the distribution, while they all seem to have trouble with the left. And finally, in Phase 4, it again seems that the 8-State model is best at reproducing the qualitative properties of the reuse distance distribution. 

The takeaway here is that in cases where the locality of the references going to the \LTWO does not matter, for instance in some phase with a very high \LDC hit rate or a phase where \LTWO and \LTHREE stats are not important to the simulation designer, it may be alright to use a model like the Fixed Hit Rate or 4-State Markov model. However, in cases where the spatial and temporal locality matters, as is typically the case, the 8-State Markov model will work best (out of the models we have to select from). 

Now that we have examined how the individual models perform, we will take a look at how we score them at simulation time. 

\begin{table}
\caption{Model Scores (lower is better)}\label{tab:scores}
\begin{tabularx}{\linewidth}{|l|X|X|X|X|X|}
\hline{}
 & Phase 0 & Phase 1 & Phase 2 & Phase 3 & Phase 4 \\ \hline\hline
Fixed Rate & 0.6875 & \textbf{0.0626} & 0.6035 & 0.9273 & 0.5188 \\ \hline
Markov 4 & 0.6519 & 0.0781 & 0.6070 & 0.7479 & 0.3842 \\ \hline
Markov 8 & \textbf{0.5802} & 0.2253 & \textbf{0.3313} & \textbf{0.6659} & \textbf{0.2668} \\ \hline
\end{tabularx}
\end{table}

\subsection{Model Selection}
\label{ssec:results:model_selection}
We described our model selection criteria in \Cref{ssec:modeleval}. In \Cref{tab:scores} we see which models actually won out, based on the criteria of accuracy, locality, complexity, and size. Each individual score is the L2-distance from the ideal score vector of $[1,1,0,0]$. Due to the ability of the 8-state Markov model to represent locality, as well as its high accuracy, we see that it wins for the majority of the phases. However, due to both the simplicity of phase 1 and the size of the Fixed Hit Rate model, the 8-state model does not win. Thus, during a run of the simulation where the simulator is  allowed to choose the best ranking model, the 8-state Markov model will be chosen for all phases except phase 1. In subsequent sections, we refer to this as \textit{ALL}, as the simulator trains all 3 models and chooses the best for each phase.  

\begin{table}
\caption{Percent Change in Cache Stats (10 Runs)}\label{tab:loc_stats}
\begin{tabularx}{\linewidth}{|l|X|X|X||X|}
\hline{}
 & L1 Hits & L2 Hits & L3 Hits & Cycles  \\ \hline\hline
Base & 7.69e+06 & 7.78e+05 & 2.53e+05 & 1.37e+08 \\ \hline\hline
Fixed Hit Rate & -0.07\% & 54.11\% & -71.11\% & -27.91\% \\ \hline
Markov 4 & -0.37\% & 46.10\% & -52.14\% & -23.13\% \\ \hline
Markov 8 & -0.19\% & 12.13\% & -4.18\% & -7.59\% \\ \hline
ALL & 0.07\% & 10.12\% & -4.65\% & -7.99\% \\ \hline
\end{tabularx}
\end{table}

\subsection{Overall Simulation Statistics}

In the last section, we saw that due to the performance of the 8-State Markov model, it is chosen by the model selection algorithm for every phase except for the marker phase, Phase 1. In this section, we'll look at how the overall simulation statistics change when using each model, including the \textit{ALL} model, as just described. This data is in \Cref{tab:loc_stats}. 

First off, we see that every model is able to accurately match the hit count of the \LDC cache. This is expected, as each model need only produce the same number of hit and misses as the base model. This is obviously the case of the Fixed Hit Rate model. This is also expected of the Markov models, as we expect them to spend the same amount of time in each state as the training data, so they are expected to produce the proper number of hits and misses. It is good, however, to know that our slightly modified Markov model, which is able to restrict the state it moves to in order to predict locality, does not break this behavior. 
As we move to look at the rest of the stats, things start looking grim for the Fixed Hit Rate and 4-State Markov models. These models do not encode any locality information, and thus the stream of misses that go to the \LTWO have far different locality properties than the base detailed model. Thankfully, it seems easy enough to correct for this. The 8-State Markov model greatly reduces the error in the \LTWO, and \LTHREE hit counts, and correspondingly the total number of execution cycles as well. 

As for the simulation where we automatically selected models, ALL, the stats are not too different from the 8-State Markov model. This is to be expected, as we saw in \Cref{ssec:results:model_selection}, the 8-State Markov model is chosen for every phase except for Phase 1, which was trivial to predict. While this choice does introduce extra error, it will be up to future simulation designers to decide if error like this is acceptable for their purposes. The fact that the Fixed Rate Model was used at all, and did not disturb overall simulation statistics too much should be seen as a win. 

Overall, our accuracy in cache hits and simulation cycles indicate that this model swapping methodology has merit, and shows promise for speeding up larger simulations.

\section{Related Work}\label{sec:related}

The topic of how to choose between models is a well studied 
topic with an established body of literature~\cite{harchol2013performance,
jain1991art,stewart2009probability}. Our work builds on these
classic works by using known modeling techniques such
as the Markov model~\cite{gagniuc2017markov}.  
A model selection survey by~\cite{piironen2017comparison} 
describes more recent methods of model selection, in
which they describe a posterior predictive
criteria method related to our distance metric. We differ in 
that we used summarized criteria in the selection 
process in addition to direct model output, as an example,
we use model complexity, and state size, in addition
to raw model performance. 

As model validation is closely related to model 
selection, our work is loosely related to 
empirical response surface evaluation~\cite{box1987empirical}.
Our work, instead of relying on complex multi-dimensional, 
time-series surfaces, uses a summary statistic over
each phase which is compared to an ideal using
a distance metric (\S~\ref{sec:onlineswapping}). 
Multiple related fields from Operations 
Research~\cite{landry1983model} and  
agriculture~\cite{gauch1988model} rely on 
model validation, typically through physical
or empirical observation. Model validation and 
simulation~\cite{sargent1979validation,sargent2005verification}
are often discussed in terms of model cost (e.g.
execution time) and model confidence (i.e., does 
the model work). We differ from the model validation
described by~\cite{sargent2005verification} in that 
due to our intended automated approach, we rely almost
exclusively on ``scoring'' vs. other more intensive
approaches. Our thesis being that we can constrain
the impact of error by constraining the modeled 
behavior by phase. 

While prior works provided the building blocks
to enable our online model swapping proof-of-concept,
we believe we are the first to demonstrate all 
such features combined, that is: phase detection,
model selection, and model swapping.


\section{Conclusions and Future Work}\label{sec:conclusions}
\begin{figure}[htp]
    \centering
    \includegraphics[width=.9\linewidth]{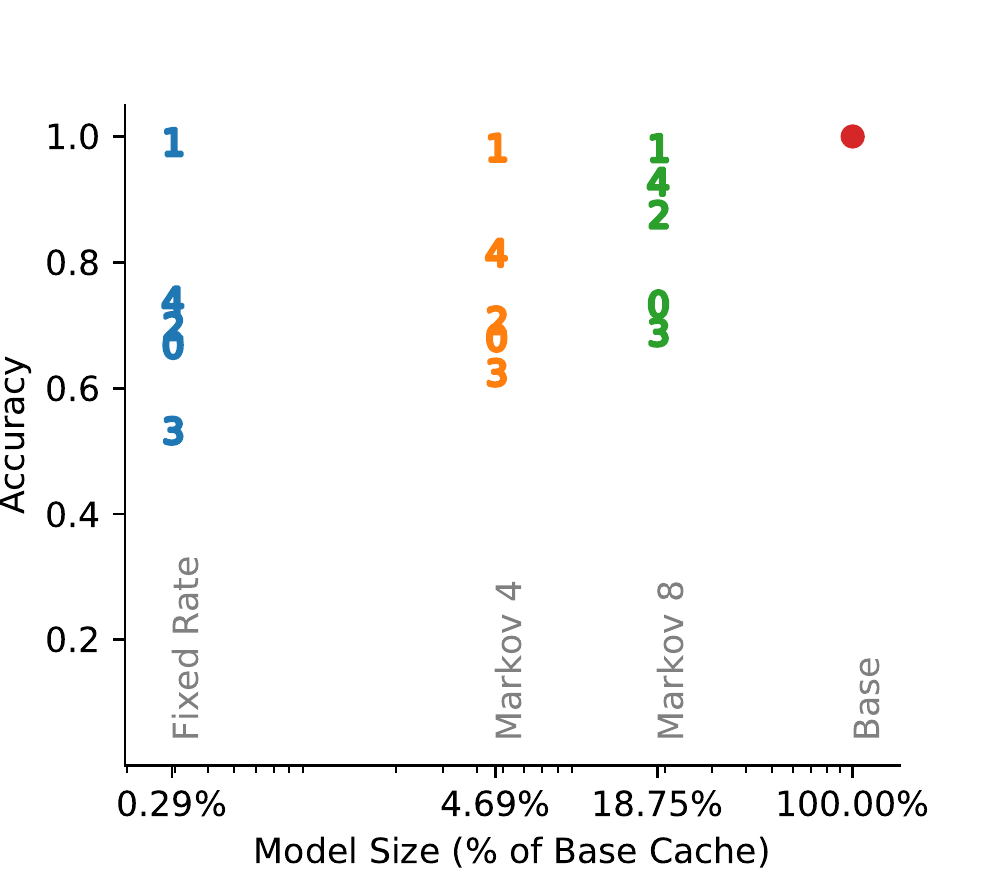}
    \caption{\textbf{Accuracy as a function of model size}: Here we visualize the trade-off between accuracy and model size, with each simpler model giving us a broader range of performance for the phases studied in this work. Our research enables simulation designers to explore this space, and other related spaces, for the purposes of faster iteration in design.}
    \label{fig:modelsize}
\end{figure}

This work demonstrates the potential for online model swapping to speed up simulation while minimizing loss in accuracy. Our proof of concept shows that simple statistical models can faithfully represent phases of execution based on the realistic kernels in Meabo. As \Cref{fig:modelsize} shows for the Meabo kernels, we can provide computer architects a new tool to quickly explore trade-offs between accuracy and the size or complexity costs of different models. We are excited by this result, but further work is yet to be done:

\begin{itemize}
\item \textbf{Integrate into SST} This project was inspired by a desire to speed up simulations using the Structural Simulation Toolkit (SST) \cite{sst}. As such, an important piece of future work will be adding a new component to SST that implements model swapping within their \textit{memHierarchy} component. This will allow our methodology to be used in more realistic simulations that include prefetching and cache coherency, and bring this work to a wide audience.

\item \textbf{More statistical models} In this work, we only have three models to choose to swap. We would like to increase the breadth of program phases that we can model well by adding more models to our simulator. In particular, we would love to add approaches from machine learning such as Recurrent Neural Nets into the mix. Although these are often thought of as quite heavy-weight, we believe that even quite simple RNNs may be able to outperform some of the models in this paper. Another avenue we would like to explore are models that have already been thoroughly evaluated, such as the models from this recent survey \cite{predictive_modeling}, provided they are fast enough to suit our needs.

\item \textbf{Model selection criteria} This work used a simple distance metric to create a ranking among models. We would like to integrate approaches from literature on model selection to strengthen our approach~\cite{claeskens2008model,anderson2003evaluating}.

In terms of the criteria we already use, it may be beneficial to use a more comprehensive metric for locality. While many locality metrics are quite computationally intensive to compute, some alternatives exist, such as the sampling based locality analysis in \cite{chen_ding,ccb19}. 

\item \textbf{More components} We used the \LDC as a starting point for us to explore model swapping. However, we designed the methodology so that it would be general enough to apply to other components in a simulation. We aim to swap out components such as other parts of the cache, parts of the network, or even pieces of the core model itself. 

\end{itemize}

The methods in this paper will serve as a starting point for future work in model swapping. With the main components of any such scheme described here, and an implementation to serve as a starting point, we know that further exploration will yield useful tools to accelerate the simulation of future compute architectures. 
\section*{Acknowledgements}

\ifthenelse{\boolean{ShowAuthors}}
{
  This work used the Hive cluster, which is supported by the National Science Foundation under grant number 1828187.  This research was supported in part through research cyberinfrastrucutre resources and services provided by the Partnership for an Advanced Computing Environment (PACE) at the Georgia Institute of Technology, Atlanta, Georgia, USA.

  Additionally, this material is based upon work supported by the National Science Foundation under Grant No. 1710371.
}{
  Acknowledgements withheld for blind review.
}

\bibliographystyle{bibliography/IEEEtran}
\bibliography{bibliography/references.bib}


\clearpage
\appendices
\section{Phase Detection}\label{appendix:phase}

A description of the phase detection algorithm that we use exists in \cite{phase_analysis}, but it does not include an implementation. We include pseudo-code here to aid the reader in understanding the implementation.

\begin{lstlisting}[label={pd_label}, caption={The pseudo-code for our phase detector implementation. On every memory reference, the IP is shared with the \texttt{phase\_detector} function, which updates a signature. At the end of an interval, this signature is compared with the previous one to see if execution is stable. },captionpos=b]
threshold    = 0.5
interval_len = 10000
sig_len      = 1024
drop_btis    = 3
sig      = new_bitvector(sig_len)
last_sig = new_bitvector(sig_len)

def diff(sig1, sig2):
    tmp1 = sig1 ^ sig2
    tmp2 = sig1 | sig2
    return popcount(tmp1) / popcount(tmp2)
    
def hash(sig):
    tmp = sig >> drop_bits
    return builtin_hash(tmp) >> (64 - log2(sig_len))
    
def phase_detector(ip):
    sig[hash(ip)] = 1 
    count += 1
    
    # If we are on an interval boundary, determine
    # the phase and notify listeners 
    if count % interval_len == 0:
        if diff(sig, last_sig) < threshold:
            stable += 1
            if stable >= stable_min and phase == -1:
                phase_table.append(sig)
                self.phase = len(phase_table) - 1
        else:
            stable = 0
            phase  = -1
            
            # Check if we have entered a phase we
            # have seen before
            if len(phase_table) > 0:
                similar = []
                for s in phase_table:
                    similar.append(diff(sig, s))
                best = imax(similar)
                if similar[best] < threshold:
                    phase = best

        last_sig = sig
        sig = new_bitvector(sig_len)
        
        notify_listeners(phase)
        
\end{lstlisting}

\section{Markov Prediction}
\label{appendix:markov}
\begin{lstlisting}[label={markov_listing}, caption={The algorithm for prediction with the 4 state Markov model. Care must be taken to only return a state that corresponds to the incoming request. This is done with the restrict parameter, which is used to restrict the states that the model can move to. In practice, the restricted prediction matrices are memoized if the \texttt{counts} matrix has not changed.},captionpos=b]

last_state  # the last state
counts      # (N x N) matrix where (i, j) is 
            # the number of times state j followed 
            # state i during training
       
# The prediction function returns the next state
# The restrict parameter is a tuple with allowed
# states
def predict(restrict):
    tmp = counts[:, restrict]
    tmp = normalize_rows(tmp)
    pred = prefixsum_rows(tmp)
    
    r = rand()
    
    for i in range(len(restrict)):
        if r < pred[last_state, i]:
            next_state = restrict[i]
            break
    
    last_state = next_state
    return next_state


\end{lstlisting}

\section{Accuracy Data}\label{appendix:accuracy_data}

\begin{table}[hp]
\caption{Mean Per-Phase Accuracy (10 Runs)}\label{tab:acc}
\begin{tabularx}{\linewidth}{|l|X|X|X|X|X|}
\hline{}
 & Phase 0 & Phase 1 & Phase 2 & Phase 3 & Phase 4 \\ \hline\hline
Fixed Rate & 0.67 & 0.99 & 0.70 & 0.53 & 0.74 \\ \hline
Markov 4 & 0.68 & 0.98 & 0.71 & 0.62 & 0.82 \\ \hline
Markov 8 & 0.73 & 0.98 & 0.88 & 0.69 & 0.93 \\ \hline
All & 0.73 & 0.99 & 0.88 & 0.69 & 0.93 \\ \hline
\end{tabularx}
\end{table}

\vspace{.25cm}


\begin{table}[hp]
\caption{Std.Dev. in Per-Phase Accuracy (10 Runs)}\label{tab:stddev}
\begin{tabularx}{\linewidth}{|l|X|X|X|X|X|}
\hline{}
 & Phase 0 & Phase 1 & Phase 2 & Phase 3 & Phase 4 \\ \hline\hline
Fixed Rate & 0.0004 & 0.0000 & 0.0002 & 0.0003 & 0.0002 \\ \hline
Markov 4 & 0.0004 & 0.0001 & 0.0002 & 0.0004 & 0.0001 \\ \hline
Markov 8 & 0.0006 & 0.0001 & 0.0001 & 0.0005 & 0.0001 \\ \hline
All & 0.0005 & 0.0000 & 0.0002 & 0.0004 & 0.0001 \\ \hline
\end{tabularx}
\end{table}


\end{document}